\documentclass{article}

\usepackage{microtype}
\usepackage{graphicx}
\usepackage{subfigure}
\usepackage{booktabs}
\usepackage{amsmath} 

\usepackage{hyperref}

\usepackage[accepted]{icml2019}

\begin{document}

\twocolumn[

\title{Improving search relevance of Azure Cognitive Search by Bayesian optimization}
\date{\vspace{-0.2in}}
\maketitle

\icmlsetsymbol{equal}{*}


\begin{icmlauthorlist}
\icmlauthor{Nitin Agarwal}{MS}
\icmlauthor{Ashish Kumar}{MS}
\icmlauthor{Kiran R}{MS}
\icmlauthor{Manish Gupta}{MS}
\icmlauthor{Laurent Bou\'e}{MS}
\end{icmlauthorlist}

\icmlaffiliation{MS}{Microsoft, CX Data Cloud + AI}

\vspace{0.4in}

\begin{abstract}

\vspace{0.4in}

Azure Cognitive Search (ACS) has emerged as a major contender in ``Search as a Service'' cloud products in recent years. However, one of the major challenges for ACS users is to improve the relevance of the search results for their specific usecases.  In this paper, we propose a novel method to find the optimal ACS configuration that maximizes search relevance for a specific usecase (product search, document search...)  The proposed solution improves key online marketplace metrics such as click through rates (CTR) by formulating the search relevance problem as hyperparameter tuning.  We have observed significant improvements in real-world search call to action (CTA) rate in multiple marketplaces by introducing optimized weights generated from the proposed approach.

\vspace{0.4cm}

\end{abstract}
\vspace{0.4in}
]

\icmlcorrespondingauthor{\\ Laurent Bou\'e}{laboue@microsoft.com}
\printAffiliationsAndNotice{}

\section{Introduction}

With over~$5.5$ million monthly active unique users (MAUs), AppSource~\cite{AppSource} is a major Microsoft web portal offering~$1^\text{st}$~party extensions as well as~$3^\text{rd}$~party SaaS solutions to Microsoft customers and partners.  Similarly, approximately~$1.1$ million MAUs go to the Azure portal (referred to sometimes in the following as the Ibiza portal) to manage and deploy their Azure resources.  Both of these web portals are crucial components of the Microsoft commercial marketplace and partner ecosystem. Considering their large number of visitors and crucial role to the overall Microsoft cloud business growth, these websites have benefited from tremendous investment in the last year with a special emphasis on foundational engineering upgrades.  Namely, one priority has been in unifying and simplifying the backend of all the web portals.  With the engineering efforts having now matured and the marketplaces’ backend foundations made more robust, emphasis has turned to the quality of the customer experience for the visitors.  Unfortunately, this has revealed a very poor state of affairs with many core services delivering sub-optimal performance.  Of particular interest is the search functionality since telemetry has revealed that almost~$500,000$ visitors make use of free-text search box in AppSource and more than~$300,000$ do so in the Ibiza portal.  

Search relevance is critical to the business success of the Microsoft marketplace as visitors expect to use these websites to discover more of the marketplace's catalog and find new SaaS products they can integrate into their own solutions.  Therefore, the question was raised:  How can one build a data-driven solution to improve the relevance of search results in response to visitors’ queries?  

Our solution, which is based on Bayesian hyperparameter tuning of Azure Cognitive Search~\cite{ACS} (ACS), delivers a significant lift in standard search relevance metrics such as normalized Discounted Cumulative Gain~\cite{Sirotkin} (nDCG).  Our methodology is generic enough that any website leveraging ACS as their search engine can apply the same technique to deliver improved search relevance specifically tuned to their own datasets.

\section{Related work}
\label{sec:relatedWork}

As search was identified as a crucial area in need of improvement, a major engineering effort was successful in unifying AppSource and Ibiza so that both websites leverage ACS as their search engine.  Unfortunately, little was accomplished in terms of customizing ACS to the specific datasets associated with AppSource/Ibiza and instead default behaviors were left in place leading to dissatisfied customers unable to find what they were searching for.  

Although Search Engine Optimization continues to be a main topic of research, we are not aware of much existing literature in this area which is specifically related to Azure Cognitive Search.  Generally speaking a few ways have been explored for ACS optimization such as:
\begin{itemize}
    \item Index size and schema: Queries run faster on smaller indexes. Periodically revisiting index composition, both schema and documents, to look for content reduction opportunities is a best practice. Schema complexity can also adversely affect indexing and query performance~\cite{ACSperf}.
    \item Query design: Number of searchable fields impacts search performance. Each additional searchable field results in more work for the search service~\cite{ACSperf}.
    \item Service capacity: Upgrading the service tier or adding replicas can increase capacity and improve performance~\cite{ACSperf}.
    \item Scoring profiles: Use weighted fields when field context is important and queries are full text search~\cite{ACSperf2}.
    \item Query boosting to give more weight to specific parts of the query~\cite{ACSperf3}.
\end{itemize}

But all of the above are heuristic and experiment-based methods.  In our proposed solution, we describe a scientific way of optimizing the weights and improving the search experience, making it reliable and robust.

\section{Search relevance optimization: Overview of the strategy}

In order to see how to optimize search relevance, it is first important to understand the mechanics of how ACS works under the hood.  

\subsection{ACS and its internal parameters}
\label{sec:underHood}

Given a corpus of content $C=\{ c_1, \cdots ,c_n \}$ (where each~$c_i$ represents a marketplace SaaS products), ACS is a cloud-based service that provides an API interface where clients (website visitors) can make requests to in order to search for specific data within the corpus.  Typically, each product listed in AppSource/Ibiza is characterized by $p$ different {\bf searchable fields} $F= \{ f_1, \cdots ,f_p \}$.  Concrete examples for those fields will be given in Section~\ref{sec:results}.

For each product $c_i$, those fields are stored by ACS into a set of different search indices so that $c_i$ can be represented as $c_i \rightarrow  c_i [f_1,… f_p]$. Once a search query $q$ is requested by a client, ACS leverages the Apache Lucene query execution engine~\cite{lucene} (query parsing, lexical analysis, document retrieval, scoring) to return a set of items from the corpus. 

The purpose of the scoring phase is to rank items so that those who are deemed to be most similar to the original search query $q$ are returned with a higher priority than others.  By default, this similarity function $r$ is defined using the well-known Okapi BM25 ranking function~\cite{bm25}.  

Each search index $f_i \in F$ is measured for similarity against~$q$ and assigned a relevance score which is a linear combination of all the individual scores evaluated from the $p$ search indices.  In other words, the similarity between the initial search query $q$ and a document $c_i \in C$ is given by 
\begin{equation}
r(q,c_i)= w_1  r \left( q, c_i [f_1 ] \right) + \cdots + w_p  r \left( q,c_i [f_p ] \right)    
\end{equation}
Sometimes to cater to the business needs we need to push products based on boosting features like popularity and transactability (paid or free) of the product. To support this ACS has a ``magnitude'' type function~$m$ which boosts products based on how high or low the numerical value is for the additional feature.  Assuming there are $t$ {\bf boosting features}~$B = \{b_1, \cdots, b_t \}$, a weighted sum of these boosting scores is assigned to the offer which is given by
\begin{equation}
m(c_i )=  u_1  m \left( c_i  [ b_1 ] \right) +  \cdots + u_t  m \left( c_i  [ b_t ] \right)
\end{equation}
Concrete examples for such boosting features will be given in Section~\ref{sec:results}.  The final relevance score of the document $c_i$ given the search query $q$ is then given by
\begin{equation}
s(q,c_i ) = m(c_i) \times r(q,c_i )   
\end{equation}

Repeating the same calculation for all items in $C$, ACS creates a ranked list whose top-20 items (adjustable) are returned to the visitor. Additionally, ACS uses so-called {\bf enhancers} $e_1$ = \texttt{ExactMatchBoost} and ~$e_2$ = \texttt{Concatenation} to provide granular instructions which impact the way search happens. The purpose of~$e_1$ is to tell~ACS about the importance to be given to the keywords in documents which match exactly as compared to those which partially match the keywords.  Finally, the purpose of~$e_2$ is to instruct~ACS whether or not to concatenate keywords in the search term if there are multiple keywords.  

\subsection{The idea: Optimizing the weights of ACS}

As mentioned above, little was done so far in the way of customizing ACS.  In particular, the set of weights~$W= \{ w_1, \cdots , w_p \, ; \, u_1, \cdots, u_t \}$ are set to some arbitrary default values.  The idea consists in using the visitors’ historical data to find what is the set of optimal weights to maximize users’ engagement with the items returned by ACS.

\subsection{Users’ historical data}
\label{sec:CTRmatrix}

\begin{figure}[ht]
\begin{center}
\centerline{\includegraphics[width=\columnwidth]{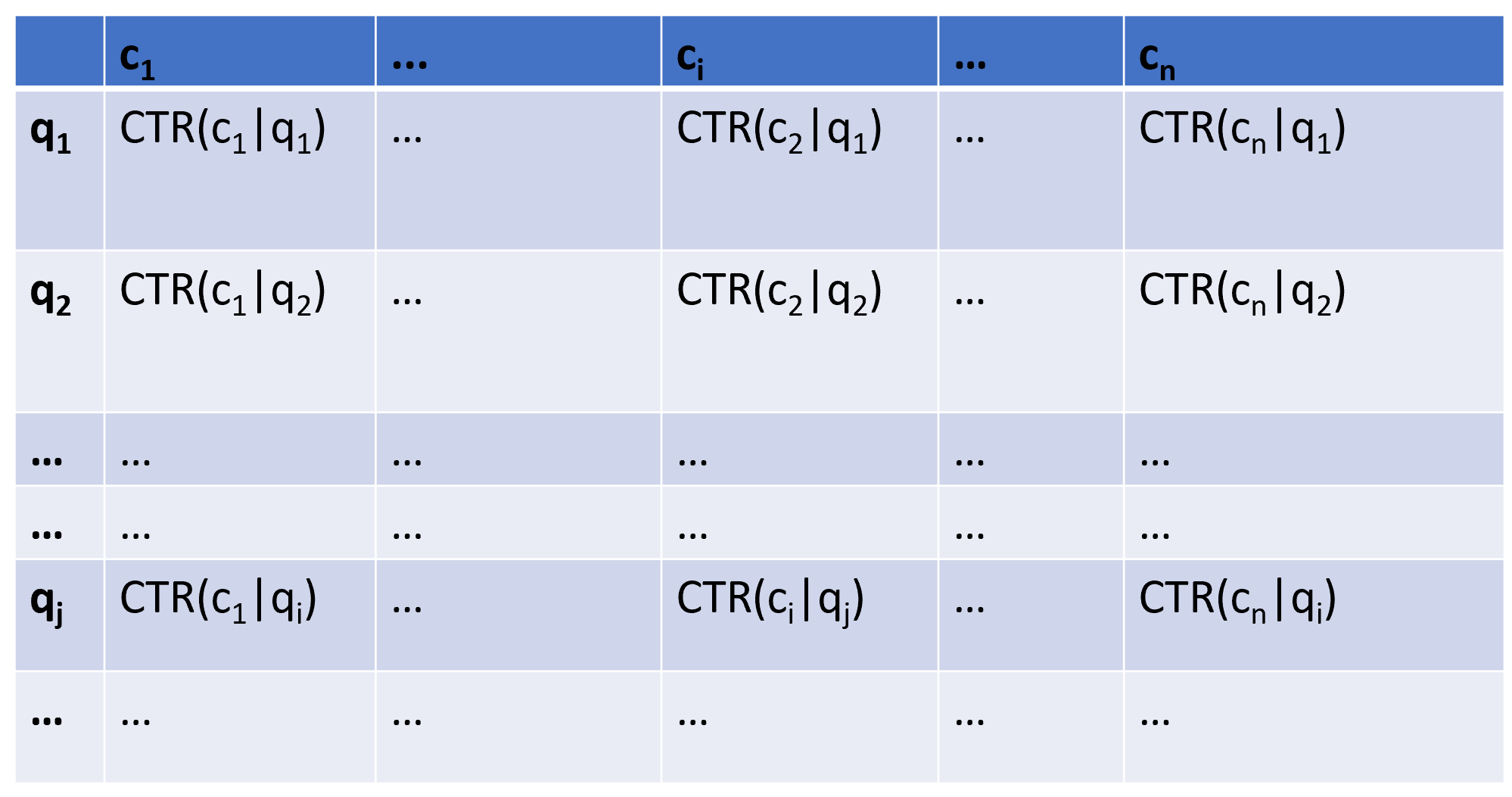}}
\caption{Illustration of the historical data.  CTR(App~$i$~$\vert$~query~$j$) represents the CTR associated with offer~$i$ conditional to the fact that the user was searching for query~$j$. }
\label{fig:ACS_data}
\end{center}
\end{figure}

In order to measure the engagement of users with the results of ACS, we have measured via web telemetry the Click Through Rate (CTR) of all items in the catalog during a search session.  For the sake of clarity, let us consider a search query~$q$ and a specific item $c_i \in C$, the proxy to estimate search relevance is measured by
\begin{equation}
\mbox{CTR} \left( c_i \, \vert \, q \right) = \dfrac{\mbox{\# clicks on } c_i \mbox{ when query is } q}{\mbox{\# impressions on } c_i \mbox{ when query is }q}
\end{equation}
Repeating this estimation of search relevance for all items in the catalog and for all observed historical search queries, we populate a very large data matrix as illustrated in Fig~\ref{fig:ACS_data} that collects our training data.

As demonstrated in Section~\ref{sec:results}, this historical CTR matrix may be populated specifically for any web portal.  This specificity is what makes the optimization procedure described in Section~\ref{sec:optimization} unique to each website that leverages ACS as its search engine.  In the following section, we will describe how this dataset is concretely used to optimize the weights~$W$.

\section{Optimizing the weights of ACS}
\label{sec:optimization}

Now that we have gathered the historical data, we can turn our attention to the optimization itself.  

Bayesian optimization~\cite{Frazier, bayesReview} is an ideal strategy for tackling the automatic optimization of non-differentiable black-box systems such as ACS whose results can only be queried via an API.  Indeed, gradients cannot be backpropagated through the API calls rendering traditional gradient-based methods inapplicable.  Although other techniques such as genetic algorithms~\cite{genetic} or iterated racing~\cite{irace} for gradient-free optimization may also be of interest in the context of ACS configuration optimization, we focus in this paper on  Bayesian optimization.  By formulating the issue of ACS configuration optimization in a manner similar to hyperparameter tuning, Bayesian optimization allows us to intelligently navigate the ACS parameter space even in the absence of derivative information.

\subsection{Defining an optimization function}

The strategy consists in looking at ACS as a black-box function which takes in two arguments: 
\begin{itemize}
    \item search query $q$
    \item set of weights $W= \{ w_1, \cdots ,w_p \, ; \, u_1, \cdots, u_t \} $
\end{itemize}
and returns 
\begin{itemize}
    \item a list of $k$ items drawn from the corpus $C$ in the order $\{c_1, \cdots ,c_k \}$ where the relevance of each item is defined as
\begin{equation}
\mbox{relevance}(c_i \,| \, q)= \mbox{CTR}(c_i \, | \, q)
\end{equation}
\end{itemize}

\noindent Then, the quality of the items returned by ACS can be measured by evaluating its normalized Discounted Cumulative Gain (nDCG) which will be defined below.

Let $\{ c_{o_1} , \cdots , c_{o_k} \}$ be the ordering of items such that
\begin{equation*}
\forall \,\, i<j \quad  \mbox{relevance}(c_{o_i} | q) > \mbox{relevance}(c_{o_j} | q)
\end{equation*}
The discounted cumulative  gain (DCG) and its corresponding ideal discounted cumulative gain (IDCG) are defined by
\begin{align*}
\mbox{DCG}(q ; W) &= \sum_{i=1}^{k} \dfrac{\mbox{relevance}(c_i|q)}{\log (i + 1)}  \\
\mbox{IDCG}(q ; W) &= \sum_{i=1}^{k} \dfrac{\mbox{relevance}(c_{o_i}|q)}{\log (i + 1)}
\end{align*}
IDCG is the DCG score for the most ideal ranking (based on historically measured relevance score such as in~Fig.\ref{fig:ACS_data}) where items are ranked top to down according to their relevance. In the equation for IDCG, $\text{relevance}(c_{o_i}|q)$ is the relevance of items from the ordered set $\{ c_{o_1} , \cdots , c_{o_k} \}$.

Finally, DCG is divided by IDCG to get the normalized discounted cumulative gain evaluated as the ratio:
\begin{equation}
\mbox{nDCG}(q ; W) = \dfrac{\mbox{DCG}(q ; W)}{\mbox{IDCG}(q ; W)}    
\end{equation}
This normalizes DCG score such that NDCG is always between 0 and 1 regardless of the length of recommendations. 

Hence, given a search query $q$, and a set of weights we have an objective function $\mbox{nDCG}(q)$ we can optimize for. Since, we have multiple search queries $Q=\{q_1, \cdots, q_z\}$ we take the weighted average of the $\mbox{nDCG}(q;W)$ values weighted by the search frequencies $\mbox{freq}(q)$ of each query~$q$. Choosing the search frequency count as the weights ensures that we give more importance to the improvement in search relevance for popular search terms. Hence, the final objective function which we want to maximize is defined as  
\begin{equation}
\mbox{nDCG}(W) = \sum_{q\in Q} f_q \times \mbox{nDCG}(q;W)
\label{eq:ndcg}
\end{equation}
where
\begin{equation}
f_q = \dfrac{\mbox{freq}(q)}{\sum_{\mbox{term}\in Q}\mbox{freq(term)}}
\label{eq:ndcg_freq}
\end{equation}

\subsection{Formulation as Bayesian hyperparameter tuning}

Since we do not have access to the internal workings of ACS, which in any case involve non-differentiable operations, we decide to optimize the set of weights using Optuna, a Bayesian Hyperparameter Optimization framework~\cite{optuna}. 

\begin{figure}[ht]
\begin{center}
\centerline{\includegraphics[width=\columnwidth]{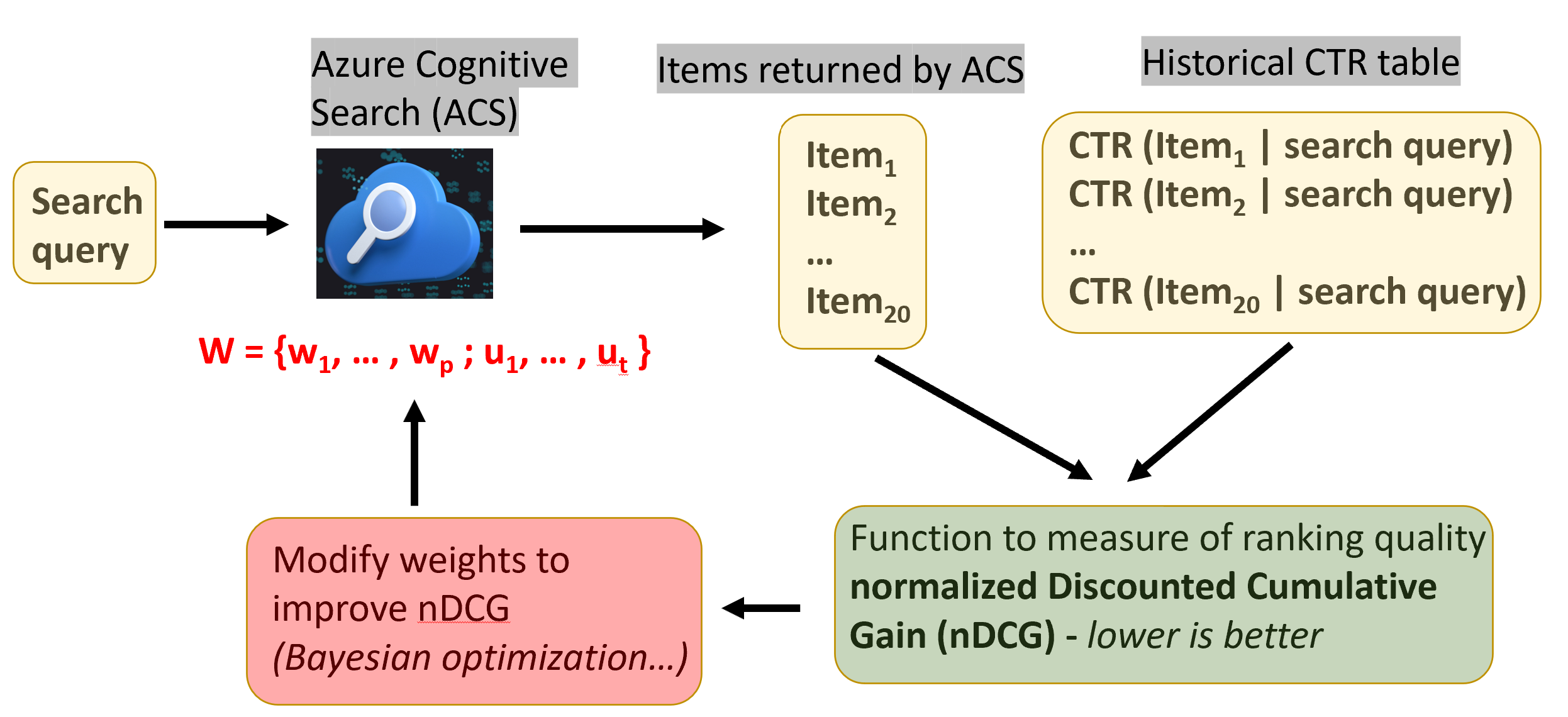}}
\caption{High level view of the model architecture.  For simplicity, we depict only a single search query as input to ACS but in practice the data processing considers larger batches of queries and the nDCG score is averaged across of the input search queries as explained in eq.~(\ref{eq:ndcg}) and eq.~(\ref{eq:ndcg_freq}).}
\label{fig:ACS_FlowDiagram}
\end{center}
\end{figure}

A description of this training loop to find the optimal values of ACS~weights~$W= \{ w_1, \cdots ,w_p \, ; \, u_1, \cdots, u_t \} $ for a specific dataset is illustrated in Fig~\ref{fig:ACS_FlowDiagram}.

Optuna finds the optimal solution using Tree-structured Parzen Estimator (TPE) and a Sequential Model-Based Optimization approach (SMBO). SMBO methods sequentially construct models to approximate the performance of hyperparameters based on historical measurements and then subsequently choose new hyperparameters to test based on this model.

In summary, we have used the weights $W=\{w_1, \cdots, w_p \, ; \, u_1, \cdots, u_t\}$ as analogous to the hyperparameters and the nDCG score as analogous to the cross-validation loss using the terminology of traditional hyperparameter tuning.  

\section{Results}
\label{sec:results}

\subsection{nDCG Performance}

We have populated the historical CTR dataset described in Section~\ref{sec:CTRmatrix} indepedently for both AppSource and Ibiza portals since they are defined by different catalogs and each has their own unique statistical properties.  For example AppSource has {$|C|_\text{AppSource} \approx$ 15,000 items in its catalog while Ibiza has~$|C|_\text{Ibiza} \approx$ 9,000}  The telemetry data was gathered directly from production data between May~$1^\text{st}$~2022 and June~$1^\text{st}$ 2022 leading to approximately {$|Q|_\text{AppSource} \approx 2,300$ distinct queries for AppSource and~$|Q|_\text{Ibiza}\approx 3,300$} queries for Ibiza.

\begin{table}[t]
\begin{center}
\begin{small}
\begin{sc}
\begin{tabular}{|l|l|l|l|}
\toprule
Field & Name in ACS & $W_\text{init}$ & $W_\text{opt}$\\
\midrule \midrule
$f_1$ & Title & 6 & 6 \\
$f_2$ & Description & 2  & 1 \\
$f_3$ & Publisher & 3 & 5 \\
$f_4$ & Categories & 2 & 4 \\
$f_5$ & Keywords & 2 & 1 \\
\midrule
$b_1$ & Popularity & 3 & 6 \\
$b_2$ & Pricing Type & 1 & 1 \\
$b_3$ & Preferred Solution & 1 & 1 \\
\midrule
$e_1$ & Exact Match & 0 & 0 \\
$e_2$ & Concatenation & 0 & 1 \\
\bottomrule
\end{tabular}
\end{sc}
\end{small}
\end{center}
\caption{Comparison of the ACS weights between their default values~$W_\text{init}$ and the predicted optimal values~$W_\text{opt}$ according to~nDCG maximization for the Ibiza portal. Within Optuna, we set the sampling distribution for each weights as uniform random integer in the range~$[1, 10 ]$.  This specific range of integers from [1-10] for the searchable~$ \{ f_1 , f_2 , f_3,f_4, f_5\} $, boosting~$ \{ b_1, b_2, b_3  \}$ and enhancer~$e_1$ fields stems from underlying ACS stakeholder requirements so our solution may be deployed in production without introducing drastic changes.  Lifting those requirements to wider intervals of floating-point values would also certainly improve even further the performance of the model.  The remaining enhancer field~$e_2$ is defined as a binary variable by construction.  Finally, the optimized weights~$W_\text{opt}$ are obtained after performing 300 trials.  The number of trials was decided after observing experimentally the convergence of~nDCG to its maximum plateau performance (not shown here).}
\label{tab:params}
\end{table}

The set of searchable~$F$ and boosting~$B$ fields (see Section~\ref{sec:underHood}) we have used for the Ibiza portal are shown explicitly in Table~\ref{tab:params}.  We are also showing the values of their associated parameters before and after their optimization.  Similar fields and numerical values have been obtained for the AppSource portal as well (not shown here).

Following the optimization procedure described in Section~\ref{sec:optimization} and using the searchable/boosting fields and enhancers listed in~Table~\ref{tab:params}, we have obtained a significant improvement as measured by nDCG in both Ibiza,~$+1.7\%$, and AppSource,~$+9.7\%$, portals as can be seen in Table~\ref{tab:results}.

\begin{table}[h]
\begin{center}
\begin{small}
\begin{sc}
\begin{tabular}{|l|l|l|l|}
\toprule
 & nDCG$_\text{init}$ & nDCG$_\text{opt}$ & nDCG lift \\
\midrule \midrule
AppSource & 0.72 & 0.79 & $+9.7\%$\\
Ibiza & 0.90 & 0.92 & $+1.7\%$\\
\bottomrule
\end{tabular}
\end{sc}
\end{small}
\end{center}
\caption{Lift in nDCG obtained after following the optimization procedure described in Section~\ref{sec:optimization}.}
\label{tab:results}
\end{table}

Accurate time-based model cross-validation is impractical due to the dynamic nature of ACS which is regularly affected by external factors (indexing updates, overall system development…) beyond our control and that may lead to inconsistencies across different time windows.  As a result, our evaluation strategy relies primarily on real-world live AB test experimentation.  During those tests, we encountered~$\approx 4,000$ distinct search terms which obviously covers a much larger scope than the top-100 search terms we used for optimizing the weights of ACS.  (Those 100 search terms represent only~$\approx 65\%$ of all search terms observed during AB tests).  With~$\approx 80,000$ searches in both treatment and control groups from March to April~2023, we have observed an improvement of about~$\approx 1.2\%$ in the CTR on the Ibiza portal (leveraging the Microsoft Experimentation Platform EXP~\cite{exp} for AB test instrumentation).


\subsection{Real-world examples of better search relevance}

Beyond the~nDCG comparison, it is also helpful to provide some real-world examples of how more relevant results are returned to the user by the optimized ACS version compared to the default ACS configuration.

\begin{figure}[ht]
\begin{center}
\centerline{\includegraphics[width=\columnwidth]{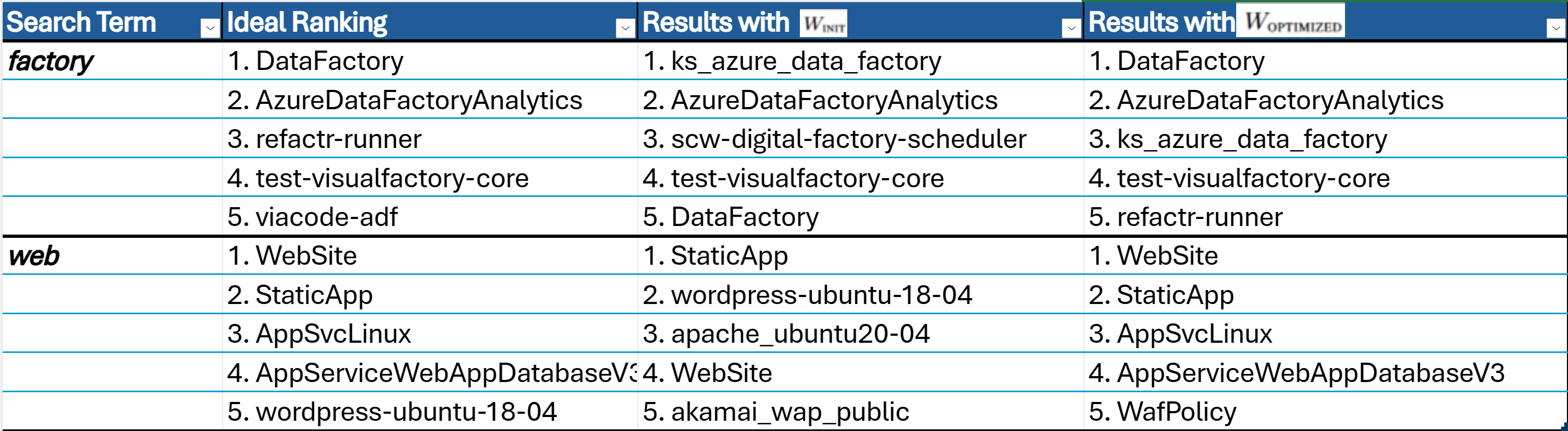}}
\caption{Illustration of the top 5 results returned by default ACS (``Current ranking'' panel), optimized ACS (``Optimized ranking'' panel) compared to the historically ideal results (``Ideal ranking'' panel) for the search terms: ``factory'' and ``web''. }
\label{fig:ACS_example}
\end{center}
\end{figure}

For example, when the query is~$q = \texttt{factory}$,~ACS should ideally respond with the item~\texttt{DataFactory} in first position.  However, using the default ACS weights ranks it only in fifth position whereas it is correctly ranked first with the optimized weights.  Another example can be seen in~Fig~\ref{fig:ACS_example}.

\section{Conclusion}

In summary, we have designed a very generic optimization system for Azure Cognitive Search. We optimize the ACS field weights to improve the overall search relevance using Bayesian hyperparameter optimization. We also find the optimal boost values for the boosting features along with the weights for the searchable fields. Our final objective function is weighted by the frequency of search terms, so that the relevance of popular terms is improved first.  The optimized weights are dataset-specific based on the telemetry data gathered in the historical CTR matrix.

In comparison to the existing set of weights, we demonstrate an uplift of 2\% in nDCG for Ibiza and 9\% in nDCG for Appsource.   We performed a UAT (User Acceptance Testing) to validate the new search results for the top 100 frequent search terms, 

Finally, a live AB test of the optimized weights vs. the default weights was gradually rolled out eventually revealing a real-world increase of the CTR by approximately~1.2\%.  Considering the amount of traffic and the importance of commercial marketplaces, this improvement was considered sufficiently satisfactory that the new optimized weights are now used by all web traffic on the portals.

Going beyond click-through rate optimization, business may require to include other aspects such as transactability, recency as part of the definition of search relevance.  This can easily be done by extending the set of searchable/boosting fields so that more parameters may also be taken into consideration in the optimization loop.

\section{Acknowledgements}

We thank our colleagues in CX Data for their feedback and support. In particular, we thank Noam Ferrara, Tal Leibovitz, Nir Levy, Boris Goldberg, Gal Horowitz, Greg Oks and Gal Keinan for their help in guiding us through the Microsoft commercial marketplaces ecosystem and partnering with us in this project.

\bibliographystyle{ieeetr}
\bibliography{MainText}

\end{document}